\documentstyle[preprint,prl,aps,epsf,epsfig]{revtex} 
\begin{document}
\draft

\title{Experimental and Theoretical Search for a Phase Transition in 
Nuclear Fragmentation}

\author{A. Chbihi$^1$, O. Schapiro$^2$, D.H.E. Gross$^2$
and S. Salou$^1$ 
}

\address{$^1$ GANIL, P.B. 5027 - 14021 Caen Cedex, France\\
$^2$ Hahn-Meitner-Institut, 14109 Berlin, Germany\\
}

\date{\today}

\maketitle

\begin{abstract}

Phase transitions of small isolated systems are signaled by the shape of the 
caloric equation of state $e^*(T)$, the relationship between the excitation
energy per nucleon $e^*$ and temperature. In this work we compare the 
experimentally deduced $e^*(T)$ to the theoretical predictions. The 
experimentally accessible temperature was extracted from evaporation spectra
from incomplete fusion reactions leading to residue nuclei.  
The experimental $e^*(T)$ dependence exhibits the characteristic S-shape
at $e^* = 2-3$ MeV/A. Such behavior is
expected for a finite system at a phase transition. 
The observed dependence agrees with predictions of the MMMC-model, 
which simulates the total accessible phase-space of fragmentation.

\end{abstract}

\pacs{25.70Np  05.70Fh 21.10Ma}

\begin{figure}
\vspace{1cm}
\begin{center}
\psfig{figure=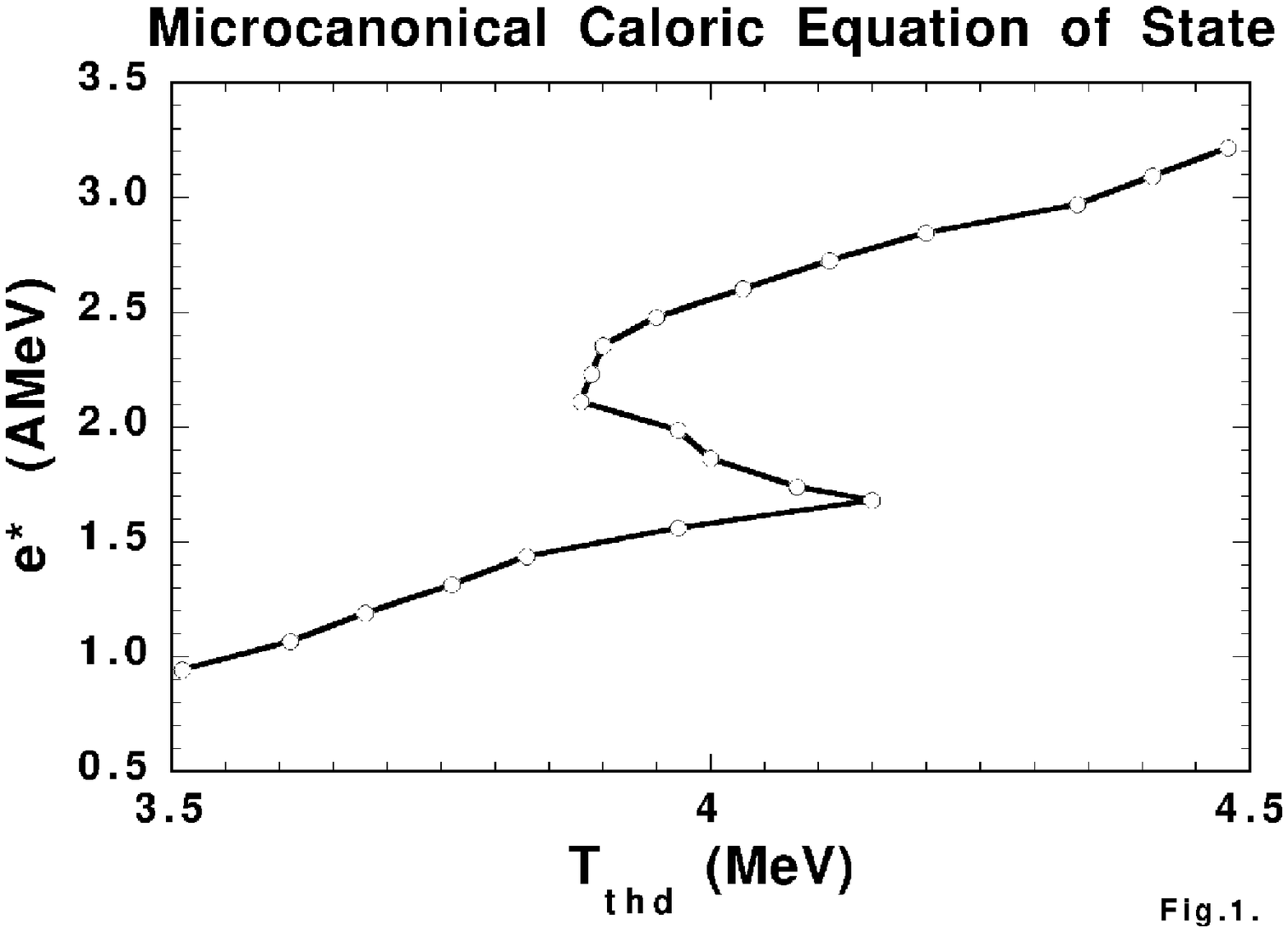,height=10cm}
\end{center}
\vspace{6mm}
\caption{Theoretical first order phase transition predicted with $M\!M\!M\!C$.
$e^*(T_{thd})$ is calculated for $A_{source}=122$, $Z_{source}=54$ and 
zero angular momentum. The computational 
error-bars are of the size of the symbol.
}
\label{fig0}
\end{figure}

In the macroscopic physics phase transitions are usually defined by
a divergence at the critical temperature, for example in heat 
capacity $c=de^*/dT_{thd}$, where $e^*$ and $T_{thd}$ is the 
excitation energy and the thermodynamic temperature. 
This is corresponding to the well known finding that at a first order 
phase transition temperature stays constant while additional energy is pumped
into the system. This picture becomes different 
if we deal with finite and isolated systems like nuclei. Due to conservation 
of mass, charge and especially total energy the signal of a first order 
phase transition is given by an "S-shape" in $e^*(T)$, called 
the caloric equation of state ($C\!E\!S$),  
as shown in the figure~\ref{fig0} for a decaying nucleus. Pictorially 
speaking we find 
that the system is cooling down with rising excitation 
energy at a first order phase transition in a finite system 
\cite{gross96,hueller94a,gross97,bixon89,mydiss}.
For a finite and isolated (microcanonical) system the heat capacity
is no longer a positive definite quantity. At first-order phase transitions
it has two divergences (instead of one for the infinite matter). In the 
region between the two poles it becomes a multi-valued function.

The signal of
fig.~\ref{fig0} can be obtained only in a microcanonical description which 
takes into account the strict  mass, charge and energy conservation.
This behavior of a fragmenting nuclear system at a phase transition  
is due to the opening of new decay channels, i.e. the population of 
additional regions of phase space $\Omega(e^*)$ 
($e^*$ is here the excitation energy per nucleon) \cite{hueller94a,gross95}.
In case of fig.~\ref{fig0} it is connected to the onset of IMF 
(intermediate mass fragment) emission and the new phase space associated to 
IMF. The specific entropy $s(e^*)=\frac{ln\Omega(e^*)}{N}$, 
N the number of nuclei,
rises then higher  compared to  a normal Fermi-gas.
It is this strong rise of entropy that leads to an anomaly in the 
$CES$ by the relation

\begin{equation}
\frac{1}{T_{thd}}\equiv\beta(e^*)=\frac{\partial s(e^*)}{\partial e^*}~~.
\label{invT}
\end{equation}

A recent experimental observation
\cite{pochodzalla95,hauger96} showing a structure in $e^*(T)$-curve 
fueled the discussion about the appearance and the measurement of 
a phase transition. It is known that apparent temperature is sensitively 
dependent on the mass of the source \cite{arnoud}. We  
suppose together with \cite{natowitz95,ma97} and in opposite to 
\cite{lee97} that the curve shown in 
ref.~\cite{pochodzalla95} is just the   
effect of changing mass of the source without undergoing any phase 
transition. Another discussion can be seen in \cite{pochodzalla97}.
Here we concentrate on a different set of experimental data deduced 
from ref. \cite{Chbihi91a}.

We are going to perform a comparison of an experimentally obtained "S-shape"
in $e^*(T)$ with theoretical predictions of the Berlin - microcanonical
statistical multifragmentation model $M\!M\!M\!C$ which simulates 
the phase space $\Omega(e^*)$ for decaying nuclei. We
are going to describe the model in some detail at a later stage. 
$M\!M\!M\!C$ predicts two phase transitions 
in nuclear fragmentation \cite{gross95}.  
The phase transition at lower excitation energy at $e^*\sim 2 - 3$ MeV per
nucleon was shown in figure~\ref{fig0}. 
A second phase transition at higher $e^*$, which is not the subject of this 
treatise and not shown here, is due to the true multifragmentation. 

Another similar statistical fragmentation model, the 
Copenhagen model SMM \cite{bondorf95} also predicts phase transitions. 
Since SMM has some mixed microcanonical-canonical features and
has a varying freeze-out volume \cite{sneppen} it 
produces a slightly different signal of a phase transition compared to  
MMMC. 

The thermodynamical temperature $T_{thd}$, equation~(\ref{invT}), 
cannot be accessed directly in an 
experiment. For the experimental comparison we need to find a related 
quantity which would keep the information on the behavior of $T_{thd}$
\cite{arnoud,gulminelli97}. 
Such a quantity, which we call apparent temperature $T_{app}$ is thus not a 
temperature in the sense of thermodynamics.

In this work we show experimentally accessible "S-shapes" of the 
$C\!E\!S$ $e^*(T_{app})$ extracted from incomplete fusion reactions 
resulting from 701~MeV $^{28}$Si + $^{100}$Mo \cite{Chbihi91a}.  
We plot $e^*$ vs. $T_{app}$, where $T_{app}$ (the apparent temperature) 
is the slope of the raw evaporation spectra.

The details of the experiment and the extraction of the needed parameters can
be found in \cite{Chbihi91a,Chbihi91}. Here we outline some of the important
features.  Heavy evaporation residues were detected at forward angles,
therefore this experiment does not probe multi-fragment final states.
Charged particles (including IMFs) and neutrons were detected in
concentric 4$\pi$-detectors. The excitation
energy of the source was deduced from linear momentum reconstruction.  The
raw spectra of protons, deuterons, tritons and alpha particles were fitted with
a three moving source prescription.  The data at backward angles are well
described by a surface-evaporating Maxwellian moving source:
\begin{equation}
\frac{d\sigma}{dE_{kin}}\propto(E_{kin}-B)\:e^{(-\frac{E_{kin}-B}{T_{app}})},
\end{equation}
where $E_{kin}$ is the center of mass kinetic energy of the particles, $B$  the
Coulomb barrier and $T_{app}$, which is the slope of the raw spectra, 
is the desired apparent temperature.

\begin{figure}
\vspace{1cm}
\leavevmode
\begin{center}
\epsfysize=11cm
\epsfbox{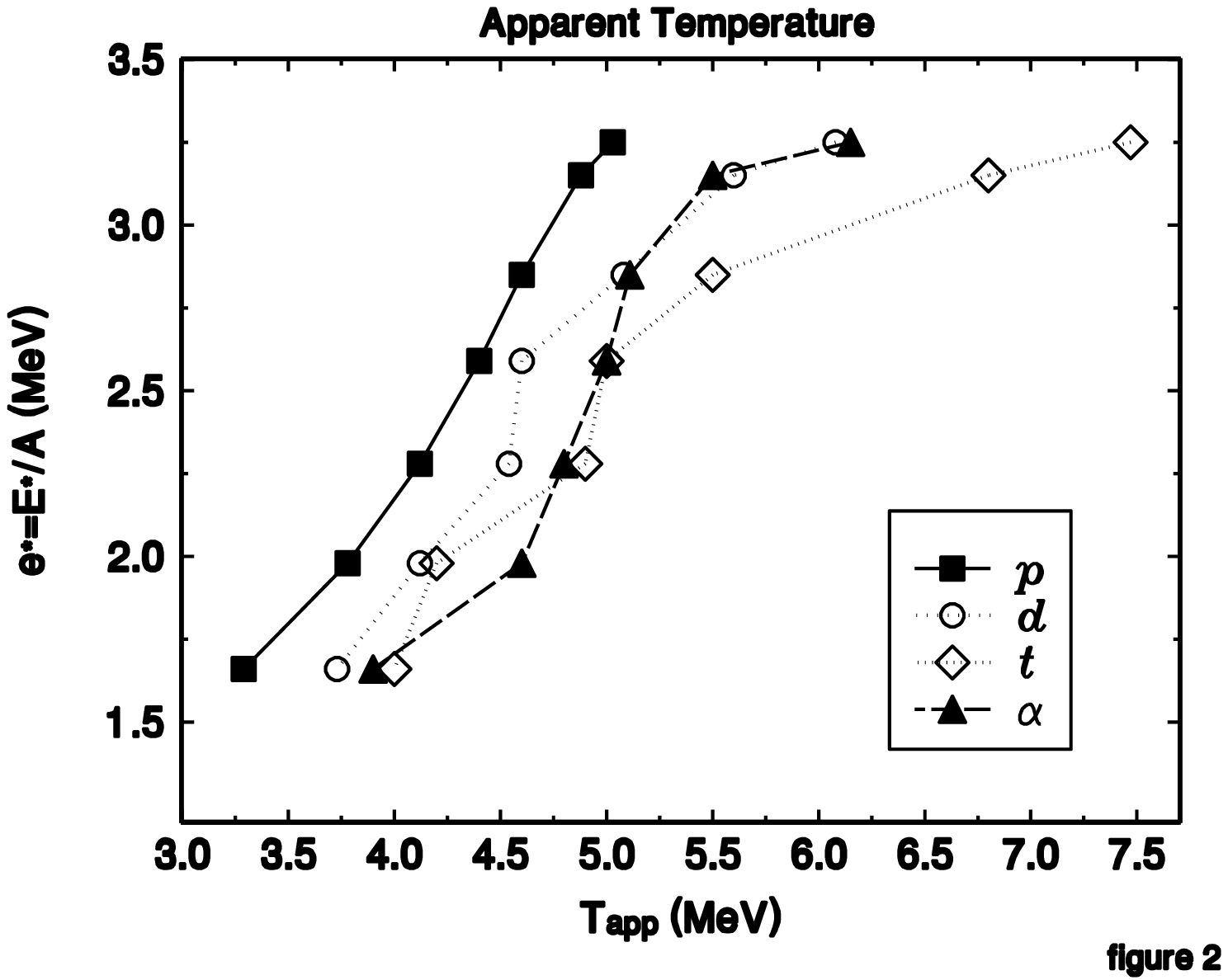}
\end{center}
\vspace{6mm}
\caption{
Experimental excitation energy per nucleon $e^*$ versus
apparent temperature $T_{app}$ for $p$, $d$, $t$ and $\alpha$.\\
The error-bars are given in fig.\ref{fig2}.}
\label{fig1}
\end{figure}

Figure~\ref{fig1} 
presents the excitation energy per nucleon $e^*$ versus $T_{app}$  
for protons, deuterons, tritons and
alpha particles.  This representation of the data exhibit two
noteworthy trends.  The first trend concerns the general shape of these curves
and the second is the horizontal displacement (along the $T_{app}$ axis) as
one progresses from protons to deuterons to tritons and alpha particles.  We
shall focus on the first observation although the second observation
is also of interest and we shall briefly discuss it also.
We find that all the four curves for different particles show an "S-shape"
in the expected region of excitation energies (compare to 
fig.~\ref{fig0}), but no backbending is seen. Here one needs to keep in mind 
that the experimental data points correspond to sources which are 
slightly changing with the excitation energy. 
From lowest to highest energy the mass is 
growing from 105 to 122 nuclei and the charge from 47 to 54. We expect this 
change in mass and charge to smear out the "S-shape" with a backbending shown 
in figure~\ref{fig0}.

Next we are going to perform a comparison to the MMMC-model simulation.

The $M\!M\!M\!C$-model assumes that the compound system fragments quite early
but the fragments remain stochastically coupled as long as they are in close
contact. Consequently, the system is equilibrated inside a freeze-out volume.
The size of this volume, which is a simulation parameter of $M\!M\!M\!C$
is in our energy region at about 6 times the normal nuclear volume. 
This corresponds to an average maximum
distance between the fragments of $\approx 2$fm. Here the nuclear
interaction between the fragments drops to the point that subsequent mass
exchange is unlikely.  Then the fragments (which can be in excited states)
leave this volume and may de-excite as they trace out
Coulomb-trajectories.  The ensuing formation of fragments  is determined by the
accessible phase space which is sampled with the
Monte Carlo method using the Metropolis importance sampling.

The experimental analysis of the data provides the values of the mass $A$,
charge $Z$, excitation energy $E^*$,  and angular momentum
$L$ of the source \cite{Chbihi91a}, which are the input into the $M\!M\!M\!C$
simulation. The only simulation parameter of the model, the 
freeze-out radius $R_f$ was taken as
its standard value of $2.2A^{1/3}$~fm, this means that we simulate a phase
transition at constant volume.  The results of $M\!M\!M\!C$
calculations, performed with these input values, were subjected to a software
filter of the experimental set-up which, most importantly, selects only those
events with one big residue.  The mass of the residue was chosen to be $A_{res}
\ge 90$, which is close to $A_{res}$ estimated from the experimental data (the
experiment did not directly measure the mass of the residue).

\begin{figure}
\vspace{1cm}
\leavevmode
\begin{center}
\epsfysize=11cm
\epsfbox{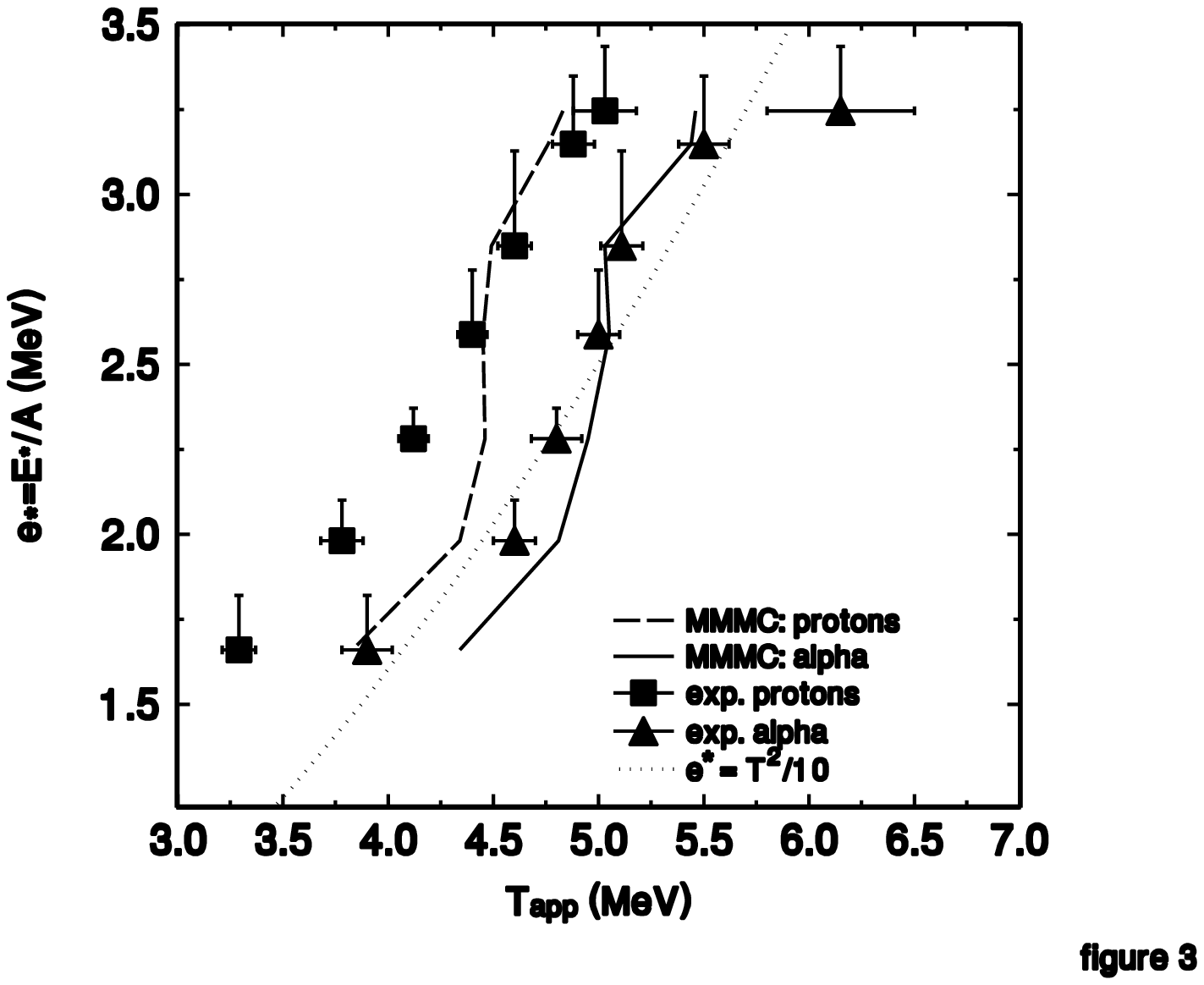}
\end{center}
\vspace{6mm}
\caption{ Experimental and theoretical (with $M\!M\!M\!C$) caloric
        equation of state, $e^*(T_{app})$ for $p$ and $\alpha$. The horizontal
error-bars give the statistical uncertainty to extract the slope from the raw 
spectra in ref.\protect\cite{Chbihi91a}. Different methods to determine the
excitation energy lead essentially to a parallel up or down shift of the curves
by the amount indicated by the vertical bars at the lowest and highest data
point. The dotted curve is a Fermi-gas calculation ($E=T^2/10$). }
\label{fig2}
\end{figure}

Figure~\ref{fig2} shows a comparison of the $e^*(T_{app})$ curves extracted from the
experimental data for protons and $\alpha$-particles to the $e^*(T_{app})$
dependence deduced from the $M\!M\!M\!C$-model \cite{gross95} using its standard
parameters. Also the experimental uncertainties for the proton and alpha curves
are given. The horizontal bars give the statistical uncertainty to extract the
temperature (slope) from the experimental raw spectra. The vertical bars (here
only given at the lowest and highest proton or alpha point) indicate the
{\em systematic} difference of the excitation energy extracted by the
''top-down'' resp.  the ''bottom-up'' procedures employed in
ref.\cite{Chbihi91a}. The two alternative methods lead essentially to an up or
down shift of the  $C\!E\!S$ curve without changing the main structure of the
curves.  The similarity of the shapes of the experimental and simulated
$C\!E\!S$ $e^*(T_{app})$ is quite evident. The differences between the shapes
of these curves and the parabolic dependence (dotted curve) expected for a
simple Fermi gas is clearly seen indicating that some additional degrees of
freedom, which are apparently included in the $M\!M\!M\!C$-model, become
significant in this energy range.

The theoretical value of $T_{app}$ was extracted from fitting, as was done for
the raw experimental spectra.  It is
worth noting that calculated temperatures $T_{app}$, extracted from 
the Maxwellian fits, are close though not identical to the unique 
thermodynamical temperatures $T_{thd}$ from the equation~\ref{invT}, as
can be seen from comparing figure~\ref{fig0} and figure~\ref{fig2}.
The curve in figure~\ref{fig0} is calculated for the mass and charge 
corresponding to the highest value of experimental energies, but for the 
whole energy range.  

The values of $R_f$ and $A_{res}$ do not influence the general shape of the
calculated $e^*(T_{app})$ curves. 
However, the $e^*(T_{app})$ curves shift along the
$T_{app}$ - axis if different values of these parameters are used.  The shifts
produced by reasonable changes in $A_{res}$ are larger than those produced by
reasonable changes in $R_f$. We checked that the anomaly in the $C\!E\!S$ is
not due to the changes of the angular momentum from $L=18.2$ to $48.8 \hbar$.
It exists also at $L=0$.

While the similarity of the shapes of the experimental and simulated
$C\!E\!S$ $e^*(T_{app})$ is good for p's and $\alpha$-particles, 
significant differences exist.  The simulated curves for deuterons and tritons
(not shown here) have the same shape but are shifted towards lower values of
$T_{app}$.  The higher $T_{app}$ values of the experimental deuteron and triton
spectra might be an indication that the production of these less bound
fragments might occur in an earlier hotter stage of the reaction, an
ingredient not included in the $M\!M\!M\!C$.

We have also compared the multiplicities of neutrons $M_n$, protons $M_p$, deuterons
$M_d$, tritons $M_t$ and alpha $M_{\alpha}$ calculated with $M\!M\!M\!C$ with
the experimental values.  The total number of the evaporated particles is the
same in the calculation and experiment. The model overestimates $M_p$ by
approximately a factor 1.3 and underestimates $M_\alpha$ by
a factor of 2. The $M\!M\!M\!C$
calculation reproduces $M_n$ and $M_t$ at all values of $e^*$.  On the other
hand, the values of $M_d$ are not reproduced by the model calculation, which
systematically predicts values which are too high. In this context one should
keep in mind: In $M\!M\!M\!C$ we treat deuterons as spherical nuclei with
normal nuclear matter density. This may overestimate their stability. 

The experimental data also suggest an association between the onset of
significant IMF production and the upswing in the $e^*(T_{app})$ dependence.
This is seen in figure~\ref{fig3} 
where the measured absolute IMF-multiplicities
($M_{IMF}$) associated with the experimentally selected events which produce a
large residue are compared to the absolute $M_{IMF}$ of the corresponding
events from the model calculation.  (The smooth dependence of $M_{IMF}$ on the
excitation energy underlines the high statistical quality of both experimental
and simulated data.)  Both the data and the calculations exhibit a dramatic
increase in $M_{IMF}$ for $e^*$ between $e^* \approx 1.5$ and $\approx 3$MeV/A.
The primary difference between the experimental data and the model predictions
is that the $M_{IMF}$'s rise at higher $e^*$-values for the model predictions
than they do for the experimental data.  Another point worth noting is that the
values of $M_{IMF}$ in these decay-channels are less than $0.1$. In other
words, most (more than $90$\%) of the selected events (both in the experiment
and in the model calculation) have no IMF.


Besides the overall agreement in the experimental an theoretical $C\!E\!S$
some uncertainty of the interpretation remains. There are too few experimental
data points outside the "S-shape" region to interpret the data unambiguously.
Further there is a small change in the charge and mass of the source for 
different excitation energies.  Therefore it is desirable to perform a similar 
experimental analysis covering a larger range of excitation energies and 
selecting a strictly constant $A_{source}$ and $Z_{source}$ for the whole 
energy range.

\begin{figure}
\vspace{1cm}
\begin{center}
\psfig{figure=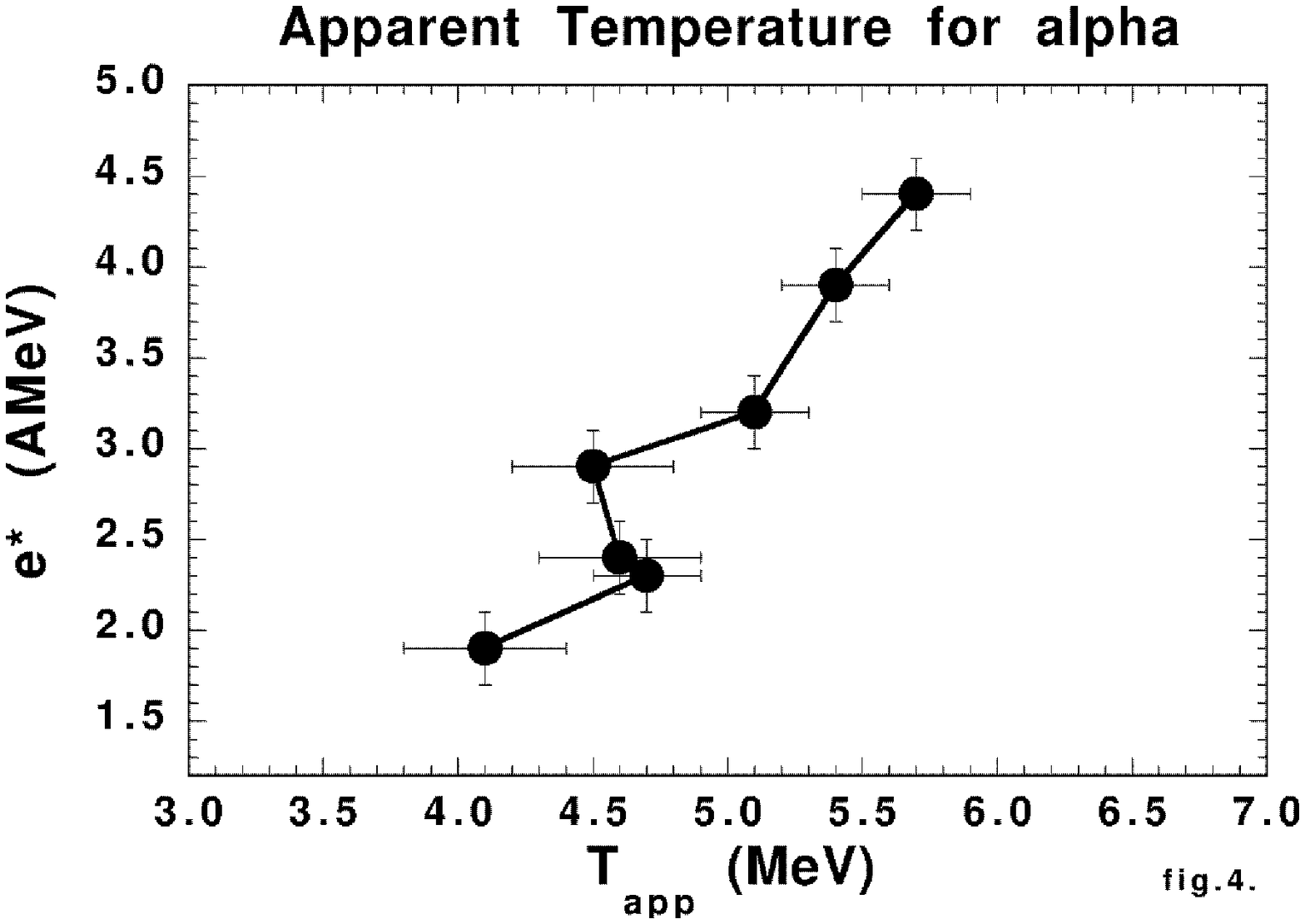,height=10cm}
\end{center}
\vspace{1cm}
\caption{The caloric equation of state for $\alpha$  using data from 
ref.~\protect\cite{wada89}.\protect}
\label{figwada}
\end{figure}

\begin{figure}
\vspace{1cm}
\begin{center}
\leavevmode
\epsfysize=12cm
\epsfbox{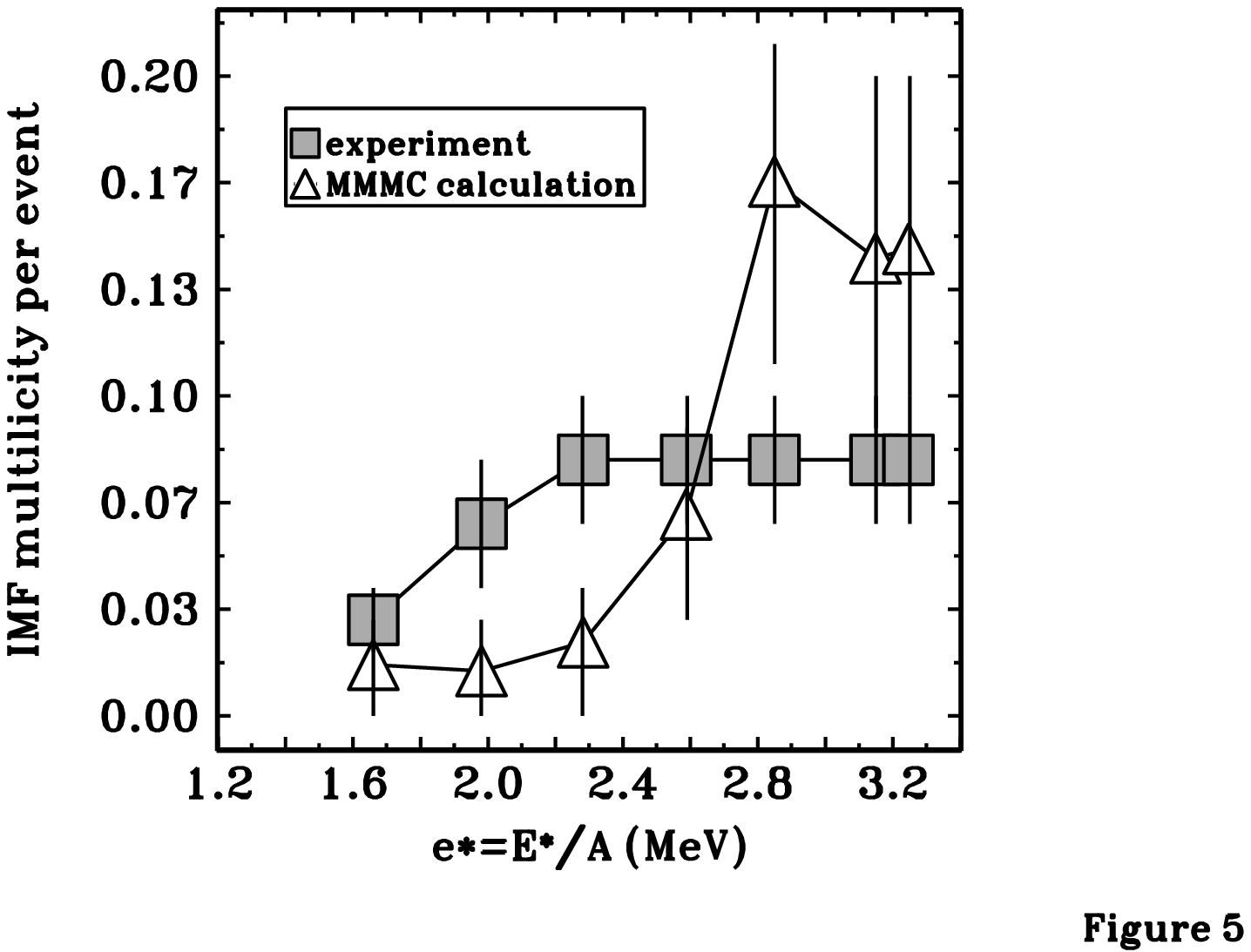}
\end{center}
\vspace{6mm}
\caption{ IMF multiplicities per event from the experiment and the
$M\!M\!M\!C$-model }
\label{fig3}
\end{figure}

Finally, we would like to point out that the extracted temperatures
$T_{app}$ for $p$ and $\alpha$-particles in related earlier experimental
work by the Texas A\&M-group \cite{wada89}, exhibit trends similar to
those presented here.  
This data is for similar masses of compound nucleus ranging from 109 to 128 
and in the same energy region. We plot the data for the apparent temperatures
$T_{app}(e^*)$ for alphas in figure~\ref{figwada}. The proton data (not
plotted here) show also a similar backbending.
Despite the large error-bars this shape anomaly  had already been 
noted and was
well reproduced by $M\!M\!M\!C$ in fig. 13 of ref.\cite{gross95}. 

~

In this paper we have shown that a strong anomaly exists in the shape of the
experimental $C\!E\!S$ $e^*(T_{app})$ for the apparent temperature. This
"S-shape" suggests that the relevant phase space becomes enlarged in the 
region of $e^* = 2-3 MeV/A$.  In terms of
thermodynamics this is  associated with a phase transition for this 
{\it isolated}, strongly interacting quantal system. 
MMMC-model reproduces the general shape of the experimental 
$e^*(T_{app})$ curve at right excitation energies.
This supports the hypothesis of strong stochastic mixing of the various 
fragmentation channels and the statistical equilibration at freeze-out. 
As the production of intermediate mass
fragments increases dramatically in this region of excitation energy we
associate the "S-shape" to the additional phase space opened by IMF production.
This association is also supported by the results of the 
$M\!M\!M\!C$ calculation where the "S-shape" in $e^*(T)$
is seen in the energy region of strongly increasing  $M_{IMF}$.

More to the point the "S-shape" in the 
caloric equation of state associated with 
IMF production is even seen in the evaporation
spectra of different particles in events which in more than $90$\% of
the cases have no IMF.  In addition, the calculations are rather insensitive to
variations of its basic model parameter, the freeze-out volume within broad
limits \cite{gross95} and thus no adjustment of this parameter was necessary to
reproduce the general shape of the experimental caloric equation of state.
Prior to this work the primary evidence of the validity of the concept of
stochastic coupling of two moving nuclei in proximity was the finding of a
strong pre-barrier surface friction in deep inelastic collisions
\cite{gross31}.
An experimental support for the equilibration hypothesis was given also 
at higher excitation energies in \cite{marie98}.

While this work adds weight to the argument that strong stochastic mixing
and equilibration exists up to rather extended configurations of the 
fragmented source, we do not believe that the issue 
is closed. For while the general shape of the caloric equation of state was
reproduced by the $M\!M\!M\!C$ model, differences exist with particle type
which may imply the existence of a dynamics or a (mean) sequence, which
is not dealt with 
by the single freeze-out configuration of the $M\!M\!M\!C$ model.

Taking all the findings together, the anomaly in all four spectra (proton,
deuteron, triton, and alpha) at the same excitation energy as predicted by
$M\!M\!M\!C$ and also the earlier data by the Texas A\&M-group 
(fig.~\ref{fig1}, \ref{fig2} and \ref{figwada}), we see a 
strong support for the significance of our interpretation of the "S-shape"
in the $e^*(T_{app})$ as a signal of a phase transition in nuclear 
fragmentation. The transition is from pure evaporation to 
asymmetric fission, which is associated to the onset of IMF emission.
Nevertheless, additional experimental and theoretical confidence is desirable.

O.S. is grateful to GANIL for the friendly atmosphere during her stays there.  
This work is supported in part by IN2P3/CNRS.






\newpage

\newpage

\newpage

\newpage

\end{document}